\documentclass[pdflatex,sn-basic]{sn-jnl}

\usepackage{lmodern}
\usepackage{amsmath,amssymb,mathtools,bm}
\usepackage{natbib}
\usepackage{graphicx}
\usepackage{overpic}
\usepackage{hyperref}
\usepackage{cleveref}
\crefname{app}{appendix}{appendices}

\newcommand{\diff}[2]{\frac{\mathrm{d}#1}{\mathrm{d}#2}}
\let\olddeg\deg
\renewcommand{\deg}[1]{\olddeg{(#1)}}

\renewcommand{\vec}[1]{\bm{#1}}

\jyear{2024}

\usepackage{lineno}

\begin{document}

\title[On discretely structured growth models and their moments]{On discretely structured growth models and their moments}

\author*[1]{\fnm{Benjamin J.} \sur{Walker}}\email{benjamin.walker@ucl.ac.uk}
\author[2]{\fnm{Helen M.} \sur{Byrne}}\email{helen.byrne@maths.ox.ac.uk}

\affil*[1]{\orgdiv{Department of Mathematics}, \orgname{University College London}, \orgaddress{\street{Gordon Street}, \city{London}, \postcode{WC1H 0AY}, \country{United Kingdom}}}
\affil[2]{\orgdiv{Wolfson Centre for Mathematical Biology, Mathematical Institute}, \orgname{University of Oxford}, \orgaddress{\street{Woodstock Road}, \city{Oxford}, \postcode{OX2 6GG}, \country{United Kingdom}}}

\abstract{The logistic equation is ubiquitous in applied mathematics as a minimal model of saturating growth. Here, we examine a broad generalisation of the logistic growth model to discretely structured populations, motivated by examples that range from the ageing of individuals in a species to immune cell exhaustion by cancerous tissue. Through exploration of a range of concrete examples and a general analysis of polynomial kinetics, we derive necessary and sufficient conditions for the dependence of the kinetics on structure to result in closed, low-dimensional moment equations that are exact. Further, we showcase how coarse-grained moment information can be used to elucidate the details of structured dynamics, with immediate potential for model selection and hypothesis testing. This paper belongs to the special collection: Problems, Progress and Perspectives in Mathematical and Computational Biology.}

\keywords{Logistic growth, structured models, moment equations}

\maketitle

\section{Introduction}\label{sec: intro}

Structure is ubiquitous in biology, from the macroscale heterogeneity of human populations down to the microscale, phenotypically structured cellular environment. Accordingly, structured models are commonly used in mathematical biology, especially those that take a discrete approach to stratification \citep{Robertson2018A20002016}. For instance, age-structured models have been used to study the spread of contagious diseases such as measles \citep{Tudor1985AnMeasles}, AIDS \citep{Griffiths2000Age-structuredEpidemic} and COVID-19 \citep{Ram2021AViruses} among different age groups, and to compare different strategies for disease control. At the cellular scale, applications include the phenotype switching of tumour-associated macrophages and the transitions of cells through the different phases of the cell cycle \citep{Eftimie2021MathematicalPhenotypes, Vittadello2019MathematicalProliferation}, along with the dynamics of lipoprotein oxidation, macrophage efferocytosis and lipid loading during early atherosclerosis \citep{Cobbold2002LipoproteinApproach,Ford2019EfferocytosisAccumulationinsidemacrophagepopulations,Ford2019APlaques,Chambers2023AMacrophages}.

A feature common to all of these discretely structured models is marked complexity. Typically, capturing structure in a discrete manner leads to a model that involves large numbers of coupled differential equations, more than might be practical to analyse in any detail without relying solely on numerical exploration. It is somewhat remarkable, then, that at least two recent works of this ilk were able to reduce their structured population models to low-dimensional systems, amenable to pen-and-paper analysis \citep{Chambers2023AMacrophages,Ford2019EfferocytosisAccumulationinsidemacrophagepopulations}. In particular, these reductions were \emph{exact moment closures}, in that the reduced system was written in terms of the moments of the population and did not introduce any additional approximations. This exactness is in stark contrast to many widespread techniques for simplifying large, possibly infinite-dimensional systems, such as asymptotic methods and approximate moment closures. In the former, one effectively takes the discrete-to-continuum limit to yield what is often a coarse-grained partial differential equation model \citep{Chambers2023AMacrophages,Wattis2020MathematicalDynamics}, which is then hoped to be more straightforward to simulate and analyse. In the latter, which is more commonly employed in active matter and stochastic applications, exact equations for the evolution of the moments of the system are formed and then truncated at some order, leading to a reduced system that is closed by explicitly relating high order moments to low order quantities \citep{Saintillan2013ActiveModels,Cintra1995OrthotropicOrientation,Feng1998ClosurePolymers,Chaubal1998AEstimation,Kuehn2016MomentReview,Gillespie2009Moment-closureModels}. Notably, no truncation was necessary in the works of \citeauthor{Chambers2023AMacrophages} and \citeauthor{Ford2019EfferocytosisAccumulationinsidemacrophagepopulations}.

The prevalence of approximate approaches for simplifying complex, highly structured models serves to highlight the apparent scarcity of exact reductions. However, in general it is not clear how often one might expect to be able to pursue an exact closure. Hence, in this article, we aim to explore and establish just how remarkable it is that the works of \citeauthor{Chambers2023AMacrophages} and \citeauthor{Ford2019EfferocytosisAccumulationinsidemacrophagepopulations} were able to identify exact moment closures in the systems that they studied. More generally, we will seek out necessary and sufficient conditions for exact closures of this type to exist in a system, though we will restrict ourselves to a particular class of structure-capturing models that focus on a single population. We postpone to future work consideration of more complex situations involving either multiple, interacting structured populations \citep{Hethcote1997AnTransmission} or populations that are structured by two or more variables \citep{Kang2020NonlinearVariables,Bernard2003AnalysisData,Restif2013QuantificationEnterica}, along with considerations of the discrete-to-continuum limit.

Seeking simplicity and some associated notion of ubiquity, we will explore population growth models that are generalisations of the classical logistic model \citep{Murray2007MathematicalIntroduction}, with the aim of laying the foundation for more elaborate future studies. Accordingly, we hope to raise more questions than we will answer. In \cref{sec: formulation}, we spend some time discussing how one might generalise logistic growth to structured populations, followed by a number of introductory and informative examples in \cref{sec: examples}. Motivated by observations and explorations from these examples, we proceed to focus on our key question, that of the existence of exact moment closures for discretely structured populations, in \cref{sec: moment closure}. In \cref{sec: example closures}, we illustrate example closures in simple yet non-trivial systems, highlighting the form and utility of exact closures and showcasing how the closure procedure can fail. Finally, we turn in \cref{sec: state-preserving vs state-resetting} to a surprising example of how studying structure can, even in populations where structure is largely absent, allow us to distinguish potential mechanisms of growth using only population-aggregated information.

\section{Model formulation}\label{sec: formulation}

\subsection{The scalar logistic equation}
For a population of size $u$, the canonical continuous-time formulation of logistic growth can be stated somewhat generally as 
\begin{equation}\label{eq: concise logistic}
    \diff{u}{t} = gu\left(1 - \frac{u}{L}\right),
\end{equation}
where $g$ and $L$ are positive parameters. Typically, $g$ plays the role of a net growth rate in a resource-rich environment, so that we might write $g = r - d$ for reproduction rate $r$ and death rate $d$\footnote{Here and throughout, we use the terms \emph{reproduction} and \emph{death} to refer to abstract processes of adding to and removing from a population, rather than to their usual definitions in specific biological contexts.}. A common alternative is to separate the processes of reproduction and death by explicitly including a death term in the model, though we do not do so here. The parameter $L$ often represents the carrying capacity of the environment, the maximum population size that is sustainable. Standard analysis of this equation, which is ubiquitous in undergraduate courses in mathematical biology, readily concludes that $u=L$ is the unique stable steady state of the differential equation for $g>0$ and it is globally stable for non-negative initial conditions \citep{Murray2007MathematicalIntroduction}.

Though the above form of the logistic equation is concise, it will be instructive to write \cref{eq: concise logistic} in the following, less concise form:
\begin{equation}\label{eq: verbose logistic}
    \diff{u}{t} = \underbrace{ru}_{\text{reproduction}} - \underbrace{du}_{\text{death}} - \underbrace{(bu)u}_{\text{burden}}.
\end{equation}
In eq. (\ref{eq: verbose logistic}) we have highlighted the different mechanisms that contribute to the logistic model. Here, the term \emph{burden} refers to the negative effects that the population has on the ability of its environment to support the population (such as the consumption of resources), with the scalar $b$ weighting this effect against the processes of reproduction and death, so that $b=g/L$ in this simplest case (where $g = r - d > 0$). In what follows, we incorporate these three mechanisms into a simple, structured model of population growth, wherein the population is stratified and there is nuance associated with distinguishing the effects of reproduction, death, and environmental burden.

\subsection{Structured populations}
As a generalisation of the classical logistic model of the previous section, we consider a population that is structured into compartments or states. In applications, a state might relate to an individual's age or to a more complex metric of health or fitness, for instance; here, we do not focus on any particular context. We structure our population discretely, indexing states by $j\in\{0,\ldots,N\}$. We write $u_j$ for the subpopulation in state $j$.

Motivated by the original logistic model, it seems natural to pose a model that is abstractly structured as
\begin{equation}\label{eq: abstractly structured logistic}
    \diff{u_j}{t} = \underbrace{R_j}_\text{reproduction} - \underbrace{D_j}_\text{death} - \underbrace{B_j}_\text{burden} + \underbrace{J_j}_\text{net flux}\,,
\end{equation}
where each undefined term is a rate that may depend on $j$ and where \emph{net flux} refers to the net rate of increase of $u_j$ due to the movement of individuals between subpopulations. One might be tempted to specify the functional forms of these effects by simply adding a general state dependence to the corresponding terms in \cref{eq: verbose logistic}. However, there is some ambiguity in taking this apparently minimal step, even when seeking the simplest possible model of growth and resource availability.

First, consider the process of reproduction. Whilst the mechanism by which new individuals are added to a population might be clear in a given context, here we will consider two plausible routes. In the first, we assume that reproduction is asymmetric and that the parent produces one offspring that is born into the lowest state, independent of the state of their parent. In this case, reproduction from any subpopulation $u_j$ contributes only to the $u_0$ compartment.

This asymmetric growth mechanism is perhaps most natural
for age-structured populations, where new individuals have minimal age at the time of reproduction. The second route is readily motivated by a simplified description of cellular growth: during mitotic division, cells produce genetic copies of themselves, which we assume inherit all accumulated genetic damage. Cast in the context of our general, structured model, this idealised, symmetric mechanism of reproduction in subpopulation $u_j$ leads to an increase in that same subpopulation. 

We seek to include contributions from both mechanisms in our model and, therefore, introduce reproduction rates per capita $r$ and $\rho$ that capture reproduction from $u_j$ into the $u_0$ and $u_j$ subpopulations, respectively. As the former mechanism effectively resets the state of the offspring whilst the latter preserves it, we will occasionally refer to these distinct mechanisms as \emph{state-resetting} reproduction and \emph{state-preserving} reproduction, respectively.

In practice, the reproductive rate in a population might be strongly linked to the state. This is often the case when structuring by fitness in mammalian populations. Therefore, we also incorporate a general state dependence into the reproductive rates. Explicitly, we take $r=r(j)$ and $\rho=\rho(j)$ to be functions of state and, to avoid ambiguity, we assume that $\rho(0)=0$, so that $r(0)$ represents the sole rate of reproduction in the $u_0$ population. With this notation and these assumptions, the reproduction term in \cref{eq: abstractly structured logistic} takes the form
\begin{equation}\label{eq: reproduction rate}
    R_j = \begin{cases}
        \sum_{i=0}^N r(i) u_i & \text{ if }j = 0\,,\\
        \rho(j) u_j & \text{ if }j > 0\,.
    \end{cases}
\end{equation}

The second term in \cref{eq: abstractly structured logistic} is not associated with the same nuance as the reproduction rate, with a linear death rate being a plausible candidate in a minimal model. Hence, taking $d(j)$ to be the state-dependent death rate per capita of the $j$\textsuperscript{th} subpopulation for some function of state $d$, we take
\begin{equation}\label{eq: death rate}
    D_j = d(j) u_j
\end{equation}
in \cref{eq: abstractly structured logistic}.

For the term that encodes the self-limiting effects of the population, $B_j$, we again adopt the principles that underpin the scalar logistic growth equation, and posit that $B_j$ depends on the overall burden of the whole population on the environment. As such, $B_j$ should inherit a dependence on the whole population, not simply $u_j$, with the nuance that subpopulations at different states might contribute differently to the overall burden. For instance, an injured animal may consume fewer resources in its environment and, thereby, be associated with a lower burden or resource cost. Hence, we write the overall burden of the population as the sum $\sum_i b(i) u_i$ for a state-dependent burden function $b$, where the sum is taken over all indices $i\in\{0,\ldots,N\}$. Further, it is plausible that different subpopulations may be differently affected by this burden on the environment. Hence, we scale the effect on subpopulation $u_j$ by a factor $e(j)$, where $e$ is an as-yet-unspecified function of state. Thus, in total, the burden term $B_j$ can be written somewhat generally as
\begin{equation}\label{eq: burden}
    B_j = e(j)\sum_{i=0}^N\left[b(i) u_i\right]u_j\,.
\end{equation}

Finally, we consider the net flux term $J_j$, which represents changes in state and, therefore, movement between compartments. Whilst there are many functional forms that $J_j$ might plausibly take, here we consider fluxes that depend linearly on the population and, hence, adopt the minimal form
\begin{equation}\label{eq: net flux}
    J_j = w(j-1)u_{j-1} - w(j) u_j
\end{equation}
for $j\in\{0,\ldots,N-1\}$, where $w$ is a state-dependent function that encodes the rate of transfer between subpopulations and is necessarily non-negative for all $j\in\{0,\ldots,N-1\}$. We extend this definition to $j=-1$ and $j=N$ by defining $w(-1)=w(N)=0$, which removes any dependence on the fictitious $u_{-1}$ and prevents flux out of the discrete state-structure domain, so that $\sum_j J_j = 0$.

The simple forms adopted here enable populations to advance in state over time, with individuals moving in the direction of increasing $j$, but they do not allow for movement in the direction of reducing $j$. Whilst this might seem like a significant and restrictive assumption, the analysis that we later pursue can be readily extended to cases where this flux is bi-directional, though at the expense of the marked brevity that uni-directional fluxes afford. Hence, we focus only on populations modelled by \cref{eq: net flux}, although we revisit bi-directional fluxes in \cref{app: both fluxes}.

In summary, our general model can be stated as
\begin{equation}\label{eq: general model}
    \diff{u_j}{t} = n(j,\vec{u})u_j + J_j + \delta_{0j}\sum_i r(i) u_i\,,
\end{equation}
where $\delta_{0j}$ is the Kronecker delta and all summations without explicit limits shall henceforth be assumed to run over $\{0,\ldots,N\}$. We have grouped the effects of state-preserving reproduction, death, and the environmental burden into a single term
\begin{equation}\label{eq: n def}
n(j,\vec{u}) = \rho(j) - d(j) - e(j)\sum_i b(i)u_i
\end{equation}
for notational convenience, writing $\vec{u}$ for the collection of all subpopulations and recalling that $\rho(0)=0$. We note that the characteristic non-linearity of logistic growth models is present in the term $n(j,\vec{u})u_j$, which includes all products of the form $u_iu_j$ though \emph{always} through the summation $\sum_i b(i)u_iu_j$. Additionally, we assume that each of the functions $r$, $\rho$, $d$, $e$, $b$ takes non-negative values on $j\in\{0,\ldots,N\}$, so that the terms in \cref{eq: abstractly structured logistic} retain their intended physical interpretation, though the analysis that follows is not sensitive to this assumption.

\section{Structured examples}\label{sec: examples}
Before we analyse the general model formulated above, it is informative to examine the consequences of structure via a number of example systems of varying complexity. These examples highlight how even simple population structures can lead to qualitative differences in overall dynamics, along with complicating pen-and-paper analysis.

\subsection{State-independent kinetics}\label{sec: examples: simple}
First, we consider one of the simplest reductions of \cref{eq: general model}, in which $r=0$, $\rho - d = \hat{g}$, $e=\hat{g}/\hat{L}$, and $b=1$, where $\hat{g}$ and $\hat{L}$ are non-negative constants that are analogous to the growth rate and carrying capacity of the scalar logistic equation of \cref{eq: concise logistic}. Put differently, this is a model in which structure plays no role in the growth kinetics. With these assumptions, the structured model becomes
\begin{align}\label{eq: simple examples: general}
    \diff{u_j}{t} &= \hat{g}u_j\left(1 - \frac{1}{\hat{L}}\sum_iu_i\right) + J_j\,,
\end{align}
which closely resembles the structure that we sought to generalise from the scalar logistic model of \cref{eq: concise logistic}, with the addition of an as-yet-unspecified flux between subpopulations. Here, in search of a direct analogue of the scalar model, we have chosen to include only state-preserving reproduction, though state-resetting reproduction will play a role in our later analysis.

Even before specifying $J_j$, we can make substantial analytical progress towards capturing the overall dynamics of this class of structured model, especially if we are only concerned with the total population. Indeed, summing \cref{eq: simple examples: general} over $j$ and defining $\mu_0=\sum_ju_j$ yields the evolution equation
\begin{equation}\label{eq: simple examples: effective logistic}
    \diff{\mu_0}{t} = \hat{g}\mu_0\left(1 - \frac{\mu_0}{\hat{L}}\right)\,,
\end{equation}
recalling that $\sum_jJ_j = 0$ and noting that the term $\sum_iu_i=\mu_0$ in
\cref{eq: simple examples: general} is independent of $j$. This evolution
equation is precisely the scalar logistic equation of \cref{eq: concise
logistic}. Thus, the overall population size of any structured population
described by \cref{eq: simple examples: general} is governed by an
unstructured logistic equation, independent of the choice of $J_j$. This
result is not unexpected, as the growth kinetics in this small class of
models are independent of the state of the subpopulation, so that moving
between subpopulations has no effect on the population's overall growth dynamics.
A key aspect of our general model formulation that allowed its reduction
to a scalar logistic equation was the assumption that the burden term for
subpopulation $j$ in \cref{eq: burden} is explicitly dependent on all
subpopulations, rather than just the $j$\textsuperscript{th} subpopulation. Indeed, if
one were to pose the evolution equation for $u_j$ with a burden that only
depends on $u_j$, such a reduction would not be possible in general.

\begin{figure}
\centering
\begin{overpic}[width=\textwidth]{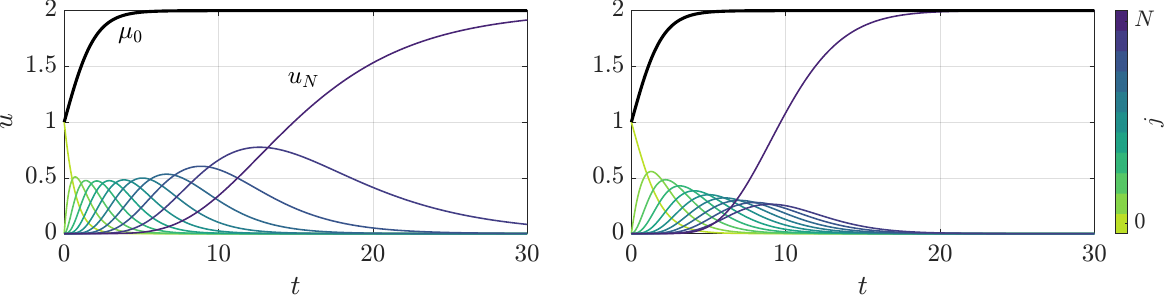}
\put(-1,25){(a)}
\put(48,25){(b)}
\end{overpic}
\caption{The dynamics of two, simple structured populations which  differ in their relations between state and flux rate. We plot the evolution of the state-structured models A and B of \cref{sec: examples: simple} in (a) and (b), respectively. In each panel, the size of the overall population is shown as a heavy black curve, whilst the dynamics of individual subpopulations are shown as lighter, coloured curves, with state indicated by colour. The dynamics of the total populations in each case are identical and logistic in character, as expected given \cref{eq: simple examples: effective logistic}, whilst the dynamics of the subpopulations differ significantly between the cases. Here, we have enforced $w(-1)=0=w(N)$ and taken $N=10$, $\hat{g}=1$, and $\hat{L}=2$, with an initial population of unit size in the $j=0$ compartment.}
\label{fig: simple examples}
    
\end{figure}

In \cref{fig: simple examples}, we illustrate the dynamics of two such models: \emph{Model A}, with $w(j) = (N - j)/W$, and \emph{Model B}, with $w(j) = 1$, where $j=0,\ldots,N-1$. The model dynamics are shown in panels (a) and (b), respectively, where $W = (N+1)/2$ fixes the average flux rate to be unity in the first model. Expectedly, both models give rise to the same overall dynamics (heavy black curves), but the dynamics of their subpopulations are markedly different (thin coloured curves). In particular, \cref{fig: simple examples}a highlights that Model A, with non-constant flux rates, prolongs the existence of subpopulations at intermediate states, commensurate with the associated decreasing flux rate. In each example, we have fixed $N=10$, $\hat{g} = 1$, and $\hat{L} = 2$, and begun the realisation with $u_0=1$ at $t=0$ and all other compartments empty.

\subsection{State-dependent models}\label{sec: examples: complex}
\begin{figure}
\centering
\begin{overpic}[width=\textwidth]{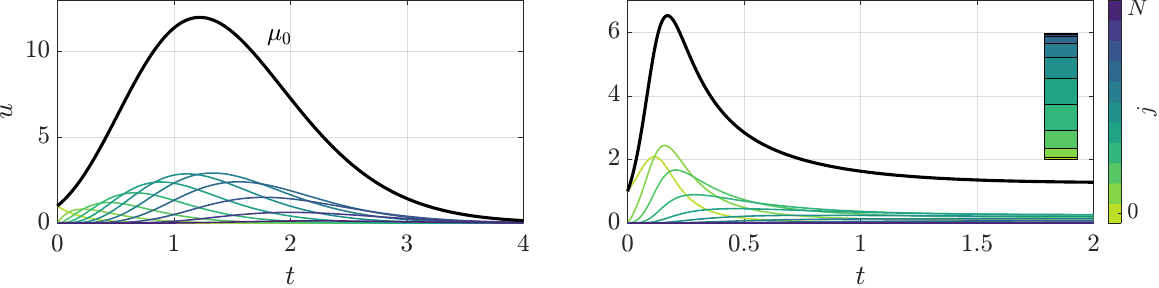}
\put(-1,25){(a)}
\put(48,25){(b)}
\end{overpic}
\caption{The dynamics of structured populations with state-dependent growth and burden, with total population sizes shown as solid black curves. The dynamics of individual subpopulations are shown as lighter, coloured curves. (a) The evolution of a model appearing to have exponential character, in which structure leads to rapid movement through compartments and eventual extinction after a period of initial growth. (b) The evolution of a more complex logistic-type model, in which growth rate declines with increasing $j$. Unlike in single-population logistic models, non-monotonicity occurs readily. Inset is a schematic of the population structure at the final timepoint, where larger regions represent larger subpopulations, highlighting a population dominated by subpopulations at intermediate states.}
\label{fig: complex examples}
\end{figure}

The simple, reduced dynamics exhibited by the previous models are far from the norm in structured population models. To illustrate this, we now consider a population with an increasing death rate $d(j) = j$, a decreasing flux $w(j) = N-j$, a constant birth rate $\beta$, and no environmental burden, so that $e(j) = 0$. This leads to the system
\begin{equation}\label{eq: complex example: model C}
    \diff{u_j}{t} = (\beta - j)u_j + J_j\,,
\end{equation}
which is exponential in character due to the lack of quadratic terms in $u_j$. Nevertheless, it demonstrates the complexity of structured models. Illustrative dynamics of this model are shown in \cref{fig: complex examples}a, exemplifying a case in which the flux rate is initially larger than the population growth rate, with $\beta=(N+1)/2$, $w(j) = N-j$, and $N=10$. With this choice of flux, the state-structured model can be written explicitly as
\begin{subequations}
\begin{align}
    \diff{u_0}{t} &= \beta u_0 - Nu_0 \,,\\
    \diff{u_j}{t} &= (\beta - j)u_j + \underbrace{[N-(j-1)]}_{w(j-1)}u_{j-1} - \underbrace{[N-j]}_{w(j)}u_j\,,\\
    \diff{u_N}{t} &= (\beta - N)u_N + u_{N-1}  \,,
\end{align}
\end{subequations} 
where $j\in\{1,\ldots,N-1\}$. This leads to a population that, despite an initial period of exponential growth, eventually collapses as the population's average state and accompanying death rate increase. In particular, the subpopulations in  compartments with small $j$ transfer rapidly into compartments of higher index and subsequently experience an increased death rate. 

As a second, truly logistic example, we include state dependence in both of the burden-related terms, taking $b(j)=j$ and $e(j)=1+(N-j)/N$. This describes a population that imposes a greater burden as they progress through states (as $j$ increases), although they also become less impacted by the overall burden. To accompany this, we include a decreasing state-preserving growth rate $\rho(j) = 2(N-j)^2/N$, with the least fit subpopulation being unable to reproduce. These assumptions lead to the more complex system
\begin{equation}\label{eq: complex example: model D}
    \diff{u_j}{t} = \left[\frac{2}{N}(N-j)^2 - \left(1 + \frac{N-j}{N}\right)\sum_i iu_i\right]u_j + J_j\,.
\end{equation}
The dynamics of this system with $w(j) = N-j$ and $N=10$ are shown in \cref{fig: complex examples}b, highlighting non-monotonic evolution towards a non-zero steady configuration. The population structure in this steady state is shown inset, highlighting that much of the population is concentrated in intermediate subpopulations. This is commensurate with the vanishing growth rate of the $j=N$ subpopulation and the rapid growth, but similarly rapid flux, of the $j=0$ subpopulation.

As the non-monotonic dynamics of the total populations in \cref{fig: complex examples} are not compatible with the monotonic solutions of the classical logistic equation, we should not expect to be able to recover such a simple model for the total population in either of these more complex cases. Focussing on the exponential model of \cref{eq: complex example: model C} and seeking an aggregated description, by summing over $j$ as before, now leads to the total population equation
\begin{equation}\label{eq: complex example: total}
    \diff{\mu_0}{t} = \beta\mu_0 - \sum_j ju_j\,.
\end{equation}
As expected, this did not eliminate the explicit dependence on each of the $u_j$, with the term $\sum_j ju_j$ remaining due to the dependence of the kinetics on the state of the subpopulation. However, the simple form of the remaining summation suggests that we seek an explicit equation for its evolution. Denoting this term by $\mu_1$ for brevity, which we remark is related to the mean state of the population by a factor of $\mu_0$, the dynamics of this weighted sum can be found by computing
\begin{equation}\label{eq: complex example: first moment}
    \diff{\mu_1}{t} = \sum_j j\diff{u_j}{t} = \beta\mu_1 - \sum_j j^2u_j + \sum_j jJ_j\,.
\end{equation}
Once again, this aggregated evolution equation contains terms that we have not yet accounted for. The pre-multiplication by $j$ before summation results in the presence of quadratic terms that cannot in general be written in terms of $\mu_0$ and $\mu_1$, whilst the details of the currently unspecified flux $J_j$ now enter into the dynamics. Without further knowledge of both the flux and the quadratic terms, further progress is not possible. In particular, whilst one might consider multiplying \cref{eq: complex example: model C} by $j^2$ instead of $j$ in an attempt to close the system, this simply generates additional unknown terms and does not result in a closed system in general. The lack of obvious closure of the aggregated system is in stark contrast to the models of \cref{sec: examples: simple}, in which closure was immediate.

At this point, it is not clear if \cref{eq: complex example: total,eq: complex example: first moment} can ever constitute a closed system, though it is plausible that some choices of the flux terms $J_j$ will lead to the necessary cancellation of terms in \cref{eq: complex example: first moment}. In the next section, through a more general analysis of \cref{eq: general model}, we will seek to determine conditions that lead to closure of these aggregated equations. As a consequence, we will in fact show that closure of this model is unattainable.

\section{Exact moment closure for polynomial state dependence}\label{sec: moment closure}
Hereafter, we consider the general model defined by \cref{eq: general model}. Generalising the approach and notation of the example of the previous section, we define the $k$\textsuperscript{th} order \emph{moment} of the population as
\begin{equation}
    \mu_k = \sum_j j^ku_j
\end{equation}
for integer $k\geq0$. Thus, $\mu_0=\sum_j u_j$ gives the total population,
$\mu_1 = \sum_j ju_j$ is the state-weighted average seen in \cref{sec:
examples: complex}, and $\mu_2$ is related to (but does not generally equal)
the variance of the state index. 

With this definition, and motivated by \cref{sec: examples: complex}, we seek to determine conditions under which the moments $\mu_0,\mu_1,\ldots,\mu_K$ form a closed description of the dynamics. We will refer to such a $K\leq N$ as the \emph{order} of a closure, noting that a system may admit closures at multiple orders. 

To make progress, we will assume that the parameters that control reproduction, the intrinsic death rate, the effect of the burden, and the flux are all polynomial functions of the state index. In terms of the notation of \cref{sec: formulation}, this implies that $r, \rho$, $d$, $e$, and $w$ are polynomials in $j$. As a result, the aggregated term $n$ is also a polynomial in $j$, noting that $\sum_i b_i u_i$ is independent of $j$ and that it can be absorbed into the coefficients of the polynomial. Explicitly, we write $n(j)$ and $w(j)$ as
\begin{equation}\label{eq: polynomial forms}
    n(j) = \sum_{\alpha = 0}^{\deg{n}}\nu_{\alpha} j^\alpha\quad \text{ and } \quad
    w(j) = \sum_{\beta = 0}^{\deg{w}}\omega_{\beta} j^\beta\,,
\end{equation}
with the coefficients $\nu_{\alpha}$ and $\omega_{\beta}$ of these polynomials being denoted by Greek characters with Greek subscripts. Here, $\deg{\cdot}$ refers to the polynomial degree, so that $\nu_{\deg{n}}$ and $\omega_{\deg{w}}$ are necessarily non-zero unless $n$ or $w$ are identically zero. We note also that the degree of $n$ is bounded above by the maximum of the degrees of $\rho$, $d$, and $e$. Here and throughout, any dependence of $n$ on $\vec{u}$ is implicit.

\subsection{The trivial closure}\label{sec: trivial closure}
Before analysing the general case, it is worth highlighting that one can \emph{always} find closed systems of moments for finite $N$. To see this, we note that the moments are simply linear combinations of the $u_j$, with coefficients $j^k$ for integer $k$. Thus, the moments and subpopulations are related via
\begin{equation}
    V^T\vec{u} = \vec{\mu}_N\,,
\end{equation}
where $\vec{\mu}_N$ is the vector of moments of order at most $N$ and $V$ is the $(N+1)\times(N+1)$ Vandermonde matrix with entries $V_{\alpha\beta} = \alpha^\beta$ for zero-based indices $\alpha,\beta\in\{0,\ldots,N\}$. It is a standard result of linear algebra that such a matrix is invertible. Hence, the moments are linearly independent and we may write each $u_j$ as a linear combination of $\mu_0,\ldots,\mu_N$ and, therefore, the same is true of all moments of order greater than $N$. Thus, we can always close a system at order $N$, which we will refer to as the \emph{trivial closure}. As this closure in general offers no reduction in complexity over the original system of equations, we will henceforth restrict our interest to closures of order $K<N$, and will neglect the trivial closure in our subsequent analysis. Moreover, we will seek closures in which $K$ is independent of $N$.

\subsection{Essentially unstructured systems}\label{sec: essentially unstructured analysis}
Consider a population in which within-subpopulation effects $n$, burden
coefficients $b$, and state-resetting growth $r$ rates are each independent of
state, so that the kinetics are independent of the structure of the
population. This can also be interpreted as a population whose governing
equations, excluding any flux terms, are invariant under all
reparameterisations of the state index $j$, a concrete example of this being the
simple systems explored in \cref{sec: examples: simple}. As we saw in
\cref{sec: examples}, these essentially unstructured systems can be reduced to
closed scalar equations for the total population by exploiting the
conservative nature of the flux, recalling that $\sum_jJ_j=0$ for any suitable
definition of $J_j$. Hence, these systems can be closed at zeroth order
without any additional constraints and always lead to dynamics of the form
\begin{equation}
    \diff{\mu_0}{t} = \left[n(\mu_0) + r\right]\mu_0\,,
\end{equation}
noting that the dependence of $n$ on $\vec{u}$ reduces precisely to a dependence on $\mu_0$ when $b$ is constant, whilst $\sum_i r(i)u_i = r\mu_0$ when $r(i)\equiv r$ is constant. 

Despite this trivial closure at zeroth order, it is not obvious if any higher
order moments can be captured quite so simply. Multiplying the evolution
equation for $u_j$ by $j^k$ and summing over the subpopulations leads to
\begin{equation}\label{eq: trivial structure: moment}
    \diff{\mu_k}{t} = n(\mu_0)\mu_k + \sum_j j^kJ_j\,,
\end{equation}
the governing equation for the moment of order $k$. Note that there is no contribution from state-resetting growth in this evolution equation, as its contribution to the sum is weighted by state index $j=0$. Significantly, the sum
$\sum_jj^kJ_j$ does not vanish in general for $k>0$, with this sum plausibly
dependent on any of the higher order moments, so that the system need not be
closed at order $k$. Hence, though we are always able to capture the evolution
of the total population via a simple logistic equation, in general we are not
able to capture finer details of the population with the same ease.

However, if $J_j$ were such that we could write $\sum_j j^kJ_j$ in terms of the moments $\mu_0$, $\mu_1$, \ldots, $\mu_K$ for every $k=1,\ldots,K$, then this system would indeed be closed, with \cref{eq: trivial structure: moment} depending on only moments of order $K$ or less. To explore whether this is possible, we insert the definition of $J_j$ from \cref{eq: net flux} into the sum for general $k$ to give
\begin{equation}
    \sum_j j^kJ_j = \sum_j j^kw(j-1)u_{j-1} - \sum_j j^kw(j) u_j\,,
\end{equation}
with sums running over $j=0,\ldots,N$. Recalling that $w(-1)=w(N)=0$, we are free to shift the index in the first summation without introducing additional terms, yielding
\begin{equation}\label{eq: weighted sum J_j}
    \sum_j j^kJ_j = \sum_j \left[(j+1)^k - j^k\right]w(j)u_j\,,
\end{equation}
valid for all admissible $w(j)$. As a polynomial in $j$, the leading order term in $(j+1)^k - j^k$ is simply $kj^{k-1}$. Hence, recalling that $w(j)$ is a polynomial in $j$ for $j\in\{0,\ldots,N\}$, the leading order term inside the summation is precisely $k\omega_{\deg{w}}j^{k-1+\deg{w}}$, the product of $kj^{k-1}$ and the leading order term of $w(j)$. 

Notably, this term vanishes for $k=0$, independent on the degree of $w$ and in line with the closure at order $K=k=0$ seen in these systems. However, this term is non-vanishing for all $k>0$, so that the contribution of the flux term to the evolution equation for $\mu_k$ contains a non-trivial dependence on $j^{k-1+\deg{w}}$ and, hence, on $\mu_{k-1+\deg{w}}$. Thus, the system of moment equations is closed at order $k$ only if $\deg{w} \leq 1$ i.e. if the flux rate $w(j)$ is constant or is linear in the state index. Somewhat remarkably, this condition is independent of $k$, so that one can find closed equations for moments of any degree in an essentially unstructured system with $\deg{w} \leq 1$. More precisely, recalling that $w$ is a non-negative polynomial and that $w(N)=0$, this class of $w$ is equivalent to non-negative scalar multiples of $N-j$. In other words, closed moment equations for such systems can be found if and only if $w\equiv C(N-j)$ for any $C\geq0$.

\subsection{General closure conditions}\label{sec: general conditions}
In essentially unstructured systems, we have seen that closure is immediate at zeroth order, and that we can sometimes obtain closed systems for higher order moments. For other systems, however, this zeroth-order closure need not occur, as we saw in the example of \cref{sec: examples: complex}. In this section, we build upon our analysis of essentially unstructured systems and explore systems that lack such simple closures.

Before we proceed, it is important to note that we can immediately place a lower bound on the order $K$ of any closure, which arises due to the form of the burden and state-resetting growth terms. In particular, as $b$ is a polynomial in $j$, the burden term $\sum_i b_i u_i$ is necessarily dependent on $\mu_{\deg{b}}$ and potentially lower order moments. Analogously, the state-resetting growth term depends on $\mu_{\deg{r}}$. Hence, these moments appear explicitly in the dynamics for each subpopulation and any closure must, therefore, include both $\mu_{\deg{b}}$ and $\mu_{\deg{r}}$. Thus, the order $K$ of any closure necessarily satisfies $K\geq\max\{\deg{b},\deg{r}\}$. Notably, this bound was trivially satisfied in our examples of essentially unstructured systems, since $\deg{b}=\deg{r}=0$ in both cases.

For the remainder of this study, we shall omit mention of these somewhat trivial lower bounds, as they do not impact consequentially on our analysis. As such, we will leave implicit any dependence of $n(j)$ on these moments, noting that they only appear as coefficients in the polynomial form of $n$ and do not otherwise complicate matters. Naturally, however, these bounds must be taken into consideration in any application of these results to a particular system.

To proceed, we return to the general form of the subpopulation dynamics given in \cref{eq: general model}:
\begin{equation}\label{eq: general model repeat}
    \diff{u_j}{t} = n(j)u_j + J_j + \delta_{0j}\sum_i r(i) u_i\,.
\end{equation}
Seeking an evolution equation for $\mu_0$, we sum over $j$ to arrive at
\begin{equation}
    \diff{\mu_0}{t} = \sum_j \left[n(j) + r(j)\right]u_j\,,
\end{equation}
noting that $\sum_jJ_j=0$. With $n(j)$ and $r(j)$ polynomials in $j$, this evolution equation can be written in terms of the moments $\mu_0,\ldots,\mu_m$, where $m = \max\{\deg{n},\deg{r}\}$. Hence, for non-constant $n$ or $r$, the dynamics of the population are not closed at zeroth order and, therefore, we must proceed to higher order moments.

Seeking the dynamics of $\mu_k$ for $k\in\mathbb{Z}^{+}$, we multiply \cref{eq: general model repeat} by $j^k$ and sum, analogously to the special case of \cref{sec: essentially unstructured analysis}. In this general case, we obtain
\begin{equation}
    \diff{\mu_k}{t} = \sum_j j^kn(j)u_j + \sum_j j^kJ_j\,,
\end{equation}
noting that the state-resetting growth term vanishes as it is weighted by $j=0$. We have already evaluated $\sum_j j^kJ_j$ in \cref{eq: weighted sum J_j}, so that this evolution equation can be written in the somewhat lengthy form
\begin{subequations}\label{eq: unidirectional moment evolution}
\begin{align}
    \diff{\mu_k}{t} &= \sum_j j^kn(j)u_j + \sum_j \left[(j+1)^k - j^k\right]w(j)u_j\\
    &= \sum_j \underbrace{\left\{j^k n(j) + \left[(j+1)^k - j^k\right]w(j)\right\}}_{p(j)}u_j\,.
\end{align}
\end{subequations}
Recalling that $n$ and $w$ are polynomials in $j$, with coefficients given in \cref{eq: polynomial forms}, the term in braces is a polynomial $p(j)$ via
\begin{equation}
    p(j) = \sum_{\alpha = 0}^{\deg{n}} \nu_{\alpha} j^{k + \alpha} + \sum_{\beta = 0}^{k-1 + \deg{w}}\phi_{k,\beta} j^{\beta}\,,
\end{equation}
where the $k$-dependent coefficients $\phi_{k,\beta}$ are defined via the convolution
\begin{equation}\label{eq: convolution}
    \phi_{k,\beta} = \sum_{\gamma_1 + \gamma_2=\beta} \omega_{\gamma_1}\binom{k}{\gamma_2}
\end{equation}
for $\gamma_1\in\{0,\ldots,\deg{w}\}$ and $\gamma_2\in\{0,\ldots,k-1\}$. For instance, we have $\phi_{k,0} = \omega_0$ and $\phi_{k,k-1+\deg{w}} = k\omega_{\deg{w}}$, whilst $\phi_{1,i}=\omega_i$ for all $i$. Hence, the governing equation for the $k$\textsuperscript{th} moment can be written explicitly as
\begin{equation}
    \diff{\mu_k}{t} = \sum_{\alpha = 0}^{\deg{n}} \nu_{\alpha} \mu_{k+\alpha} + \sum_{\beta = 0}^{k-1 + \deg{w}} \phi_{k,\beta}\mu_{\beta}\,,
\end{equation}
with powers of $j$ in $p(j)$ giving rise to the corresponding moment after summation.

This explicit form reveals an expected complication: if the growth kinetics depend on state, so that $\deg{n}>0$, the evolution equation for $\mu_k$ contains a dependence on $\mu_{k+1},\ldots,\mu_{k+\deg{n}}$. An analogous property emerges from the flux terms: unless their dependence on state is linear or trivial, they contribute moments $\mu_{k+1},\ldots,\mu_{k-1+\deg{w}}$. Thus, the only way for closure to be possible at order $K=k$ is if the contributions of these terms precisely cancel out for all moments of order greater than $k$. In particular, this must be true for the highest order moments, so that
\begin{equation}
    \nu_{\deg{n}}\mu_{K+\deg{n}} + K\omega_{\deg{w}}\mu_{K-1+\deg{w}} \equiv 0\,.
\end{equation}
As the moments are linearly independent (as established in \cref{sec: trivial closure}) and the leading order coefficients of the polynomials are non-zero by definition, this equality can only hold if the moments are in fact the same and their coefficients sum to zero. In symbols, this imposes the degree constraint
\begin{equation}\label{eq: degree constraint}
    \deg{n} = \deg{w} - 1
\end{equation}
along with a coefficient constraint 
\begin{equation}\label{eq: coefficient constraint}
    \nu_{\deg{n}} + K\omega_{\deg{w}} = 0\,.
\end{equation}
Remarkably, \cref{eq: degree constraint} is independent of $K$ and is therefore necessary for closure at \emph{any} order. In other words, closure is only possible if the polynomial degree of the flux rate is one more than the degree of the kinetic terms; otherwise, closure is not achievable.

Assuming that this degree condition is satisfied, the complete set of conditions on the coefficients needed to eliminate any explicit dependence on moments $\mu_{K+1},\ldots,\mu_{K+\deg{n}}$ may be simply stated as
\begin{equation}\label{eq: general coefficient constraint}
    \nu_{\alpha} + \phi_{K,K+\alpha} = 0
\end{equation}
for $\alpha\in\{1,\ldots,\deg{n}\}$, a linear system of equations relating the growth kinetics to the transport. Note that the number of conditions depends only on $\deg{n}=\deg{w}-1$ and not on the order of the closure. However, the conditions themselves depend on $K$, so that closure at different orders imposes different constraints on the forms of $n$ and $w$. Therefore, a system that is closed at order $K$ should not in general be expected to be closed at any other order, as this additionally requires $\phi_{K,K+\alpha} = \phi_{l,l+\alpha}$ for some $l\neq K$ and all $\alpha$.

This observation proves to be restrictive beyond just the uniqueness of $K$ for a given system. At this point, it may appear that closure is always possible for any given $n$: the degree of $n$ fixes the degree of $w$, whose coefficients we then partly specify by solving a system of the form of \cref{eq: general coefficient constraint}. However, we note that these conditions arise only as the result of requiring that the evolution equation for the $K$\textsuperscript{th} moment does not contain higher order moments, and provide no such guarantees for the other evolution equations.

To see the impact of this, note that the moment $\mu_{k+\deg{n}}$ will appear in the evolution equation for $\mu_k$ if it is not removed by an appropriate flux term. For $k=K$, the above conditions on $\deg{n}$, $\deg{w}$, and the coefficients of $w$ guarantee that introducing an appropriate flux will eliminate this term and other moments of order greater than $K$ from the evolution equation for $\mu_K$. Thus, the system appears to be closed at order $K$. If $\deg{n}=1$, then the system truly is closed: for each $k<K$, the evolution of $\mu_k$ will depend on $\mu_{k+1}$, but the system closes overall as the evolution of $\mu_K$ depends only on lower order moments. Hence, the conditions established above are sufficient for guaranteeing closure when $\deg{n}=1$.

However, for $\deg{n}>1$, the provided conditions are \emph{not} sufficient. In this case, the evolution equation for $\mu_{K-1}$ will depend on $\mu_{K-1+\deg{n}}\not\in\{\mu_0,\ldots,\mu_K\}$ and need not be cancelled out by a corresponding flux term. In fact, such cancellation can \emph{never} occur. To see this, note that we necessarily have
\begin{equation}\label{eq: subleading moment problem}
    \diff{\mu_{K-1}}{t} = \left[\nu_{\deg{n}} + \phi_{K-1,K-1+\deg{n}}\right]\mu_{K-1+\deg{n}} + \cdots
\end{equation}
where terms of lower degree have been omitted. Moreover, $\phi_{K-1,K-1+\deg{n}} = (K-1)\omega_{\deg{w}}$ from \cref{eq: convolution}, so that the dependence on $\mu_{K-1+\deg{N}}$ vanishes if and only if
\begin{equation}\label{eq: subleading moment problem: coefficient constraint}
    \nu_{\deg{n}} + (K-1)\omega_{\deg{w}} = 0\,.
\end{equation}
Though this seems like a straightforward condition to impose, we note that \cref{eq: coefficient constraint} already fixes the leading coefficient $\omega_{\deg{w}}$ of $w$, so that \cref{eq: subleading moment problem: coefficient constraint} becomes
\begin{equation}
    \nu_{\deg{n}} = 0\,.
\end{equation}
This is a contradiction as $\nu_{\deg{n}}$ is nonzero by definition. Hence, closure is not possible if $\deg{n}>1$, irrespective of the choice of flux or the precise form of the growth dynamics. We will illustrate this failure in the final example of \cref{sec: example closures}.

Strictly, so far we have only established that the derived conditions are necessary for closure if we restrict to closures that are constructed via the above procedure. It remains to show that these conditions preclude the existence of \emph{any} closure. In pursuit of this, suppose that we have failed to satisfy the conditions derived above, leading to an evolution equation of the form
\begin{equation}
    \diff{\mu_k}{t} = g(\mu_0,\ldots,\mu_N) = g(\vec{\mu}_N), 
\end{equation}
where the function $g$ depends non-trivially on at least one $\mu_i$ for $i>K$. Seeking a contradiction, suppose also that a closure does exist at order $K<N$. Necessarily, we must be able to write $g(\vec{\mu}_N) = h(\vec{\mu}_K)$ for some function $h$. However, in order for this relation to hold in general, it must hold for all initial values of $\vec{u}$. As the relationship between $\vec{u}$ and $\vec{\mu}_N$ is invertible from \cref{sec: trivial closure}, arbitrary initial conditions for $\vec{u}$ translate into arbitrary initial conditions for the moments $\vec{\mu}_N$. Hence, the equality between $g$ and $h$ must hold for all arguments, which is a contradiction as precisely one of them depends non-trivially on at least one $\mu_i$ for $i>K$. Hence, no such $h$ exists and our conditions are both necessary and sufficient for generating non-trivial closures.

In summary, we have shown that structured logistic systems with growth kinetics that are linear in the state index $j$ can be closed exactly at any order $K$ by constraining the flux term via \cref{eq: coefficient constraint,eq: degree constraint}. In \cref{sec: example closures in structured systems}, we will present an example that explicitly gives the general form of admissible fluxes. For all other growth dynamics, i.e. those with explicit nonlinear dependence on the state $j$, we have shown that non-trivial moment closure of this form is impossible.

\section{Example closures}\label{sec: example closures}
In this section, we present examples of closures in essentially unstructured and structured populations, along with an example in which closure is not possible due to nonlinear growth kinetics.

\subsection{Essentially unstructured systems}\label{sec: examples in unstructured systems}
\begin{figure}
    \centering
    \begin{overpic}[width=\textwidth]{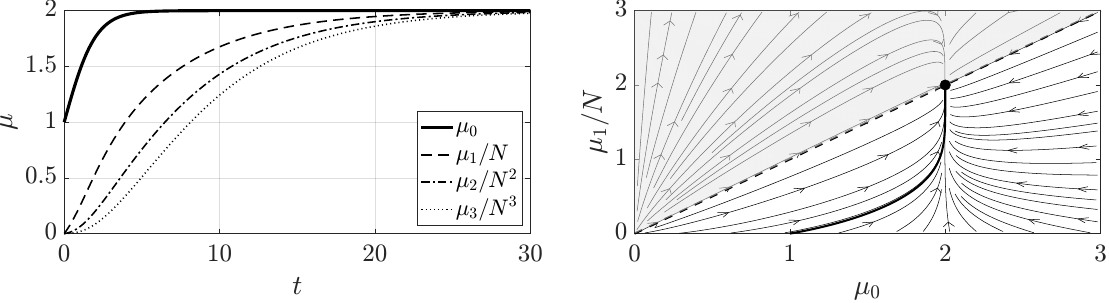}
    \put(-1,27){(a)}
    \put(50,27){(b)}
    \end{overpic}
    \caption{The evolving moments of Model A. (a) Using the same parameters and initial conditions as \cref{fig: simple examples}, we plot the evolution of the lowest order moments corresponding to Model A, an essentially unstructured system that admits closures at every order. Moments have been rescaled by powers of $N$ for visual clarity. In (b), we plot the phase portrait of the dynamics in the $\mu_0$-$\mu_1$ plane, scaling $\mu_1$ by $N$. As the system closes at this order, the phase diagram completely captures the evolution of the lowest moments. The particular case shown in (a) is plotted here as a thick black curve, with the population growing quickly overall and then being transported to subpopulations with larger $j$. The upper, shaded region is infeasible, bounded by $\mu_1 = N\mu_0$, and the black disc is the unique stable steady state.}
    \label{fig: moments of model A}
\end{figure}

Both of the examples considered in \cref{sec: examples: simple} are
essentially unstructured. Indeed, Model A has a flux rate that is precisely of
the form $w\equiv C(N-j)$, with $C=1/W$. Computing the
evolution equations for the first- and second-order moments yields
\begin{align}
    \diff{\mu_1}{t} &= \hat{g}\mu_1\left(1 - \frac{\mu_0}{\hat{K}}\right) + \frac{N\mu_0 - \mu_1}{W}\,, \\
    \diff{\mu_2}{t} &= \hat{g}\mu_2\left(1 - \frac{\mu_0}{\hat{K}}\right) + \frac{N\mu_0 + (2N-1)\mu_1 - 2\mu_2}{W}\,.
\end{align}
As expected, these constitute a closed system of moment equations for $\{\mu_0,\mu_1,\mu_2\}$ when combined with \cref{eq: simple examples: effective logistic}, along with the closed subsystems $\{\mu_0,\mu_1\}$ and even $\{\mu_0\}$. Hence, we have explicit closures at orders $K\in\{0,1,2\}$ for this system, and the above analysis predicts similar closures for all $K$. We illustrate the evolution of the first three moments in \cref{fig: moments of model A}a, along with a phase portrait of the closed dynamics of $\mu_0$ and $\mu_1$ in \cref{fig: moments of model A}b. 

Model B of \cref{sec: examples: simple}, however, is qualitatively different: as the flux rate is constant and nonzero for $j\neq N$, a polynomial of degree at least $N$ is required to interpolate it. Hence, even the evolution equation governing its first-order moment $\mu_1$ will contain a dependence on a moment of order $N$ or higher, so that the only possible closures are that at order $K=0$ or the trivial closure of \cref{sec: trivial closure}. Thus, perhaps surprisingly, apparent simplicity is not sufficient to yield non-trivial low-order closures in structured growth models.

\subsection{Structured systems}\label{sec: example closures in structured systems}
We now illustrate the general results formulated in \cref{sec: general conditions} by returning to the two structured examples of \cref{sec: examples: complex}, which do not close at zeroth order. Rather than analyse a model with a specified flux rate $w$, we will instead determine the flux rates that are compatible with closure of these models at various orders.

\subsubsection{Structured exponential growth}
We begin with the exponential model of \cref{eq: complex example: model C}, illustrated in \cref{fig: complex examples}a, whose dynamics we recount as
\begin{equation}
    \diff{u_j}{t} = (\beta - j)u_j + J_j\,.
\end{equation}
Here, we have $n(j)=\beta-j$ and $r(j)=0$, so that any closures must be of order $K\geq\deg{n}=1$. As $\deg{n}=1$, the analysis of \cref{sec: general conditions} guarantees that closure is possible, and specifies $\deg{w}=\deg{n}+1=2$, so that the only fluxes that permit closures are (at most) quadratic in the state index $j$. The coefficients of $w$ are then determined (in part) by \cref{eq: coefficient constraint}, which in this case reduces to
\begin{equation}
    \omega_2 = -\frac{\nu_1}{K} = \frac{1}{K}\,,
\end{equation}
as $\nu_1 = -1$ and $\phi_{K,K+1} = K\omega_2$. Hence, as the conditions of \cref{sec: general conditions} are necessary and sufficient for closure when $\deg{n}=1$, this system can be closed for any flux rate of the form
\begin{equation}
    w(j) = \frac{1}{K}j^2 + \omega_1 j + \omega_0\,.
\end{equation}
Recalling that $w(N)=0$ and $w(j)\geq0$ for $j\in\{0,\ldots,N\}$, the admissible fluxes reduce to a one-parameter family of polynomials of the form
\begin{equation}
    w(j) = \frac{1}{K}(j-N)(j-C)
\end{equation}
for all $C\geq N$. More generally, linear growth kinetics may be closed by fluxes of the form
\begin{equation}
    w(j) = \frac{\nu_1}{K}(j-N)(C-j)\,,
\end{equation}
where $C \geq N$ if $\nu_1<0$ and $C \leq 0$ if $\nu_1>0$, recalling that $w(j)\geq0$ for admissible $j$. Somewhat curiously, the leading factor of $1/K$ implies that closure at higher orders requires reduced fluxes. 

\begin{figure}
    \centering
    \begin{overpic}[width=\textwidth]{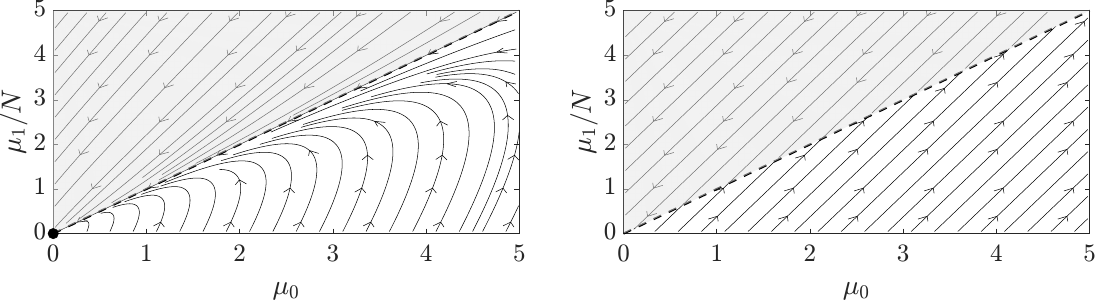}
    \put(-1,27){(a)}
    \put(50,27){(b)}
    \end{overpic}
    \caption{Phase portraits of a structured exponential model. Using the closed moment equations of \cref{eq: exponential moments}, we illustrate the phase portraits governing the evolution of $\mu_0$ and $\mu_1$. In (a), we choose $\beta = N/2 < N$, so that $(0,0)$ is linearly stable. In (b), we take $\beta=N$, leading to a continuum of marginally stable steady states on the curve $\mu_1=N\mu_0$. Here, $N=10$ and $C=2N$. The upper, shaded region in both panels is infeasible, bounded by $\mu_1 = N\mu_0$, shown dashed.}
    \label{fig: exponential phase portraits}
\end{figure}

As validation of our closure-predicting analysis, we explicitly compute the closure at order $K=1$, giving
\begin{subequations}\label{eq: exponential moments}
\begin{align}
    \diff{\mu_0}{t} &= \beta\mu_0 - \mu_1\,,\\
    \diff{\mu_1}{t} &= CN\mu_0 + (\beta-C-N)\mu_1\,,
\end{align}
\end{subequations}
noting that the $\mu_2$-term generated by $n$ in the equation for $\mu_1$ has been precisely cancelled out by the flux, as expected. This system, whose dimension and complexity is independent of $N$, is simple to study. The extinction state $(\mu_0,\mu_1)=(0,0)$ is linearly stable if and only if $\beta < N$, and gives way to a marginally stable continuum of fixed points for $\beta = N$. Larger $\beta$ correspond to rapid exponential growth with no admissible non-zero steady states. Two phase portraits are shown in \cref{fig: exponential phase portraits}, with $\beta<N$ in \cref{fig: exponential phase portraits}a and $\beta=N$ in \cref{fig: exponential phase portraits}b, with the latter highlighting the continuum of steady states with $\mu_1 = N\mu_0$ when $\beta=N$.

Before we move on to a final example, it is worth noting that care should be taken when studying the closed system of moments, as they may capture more behaviours than are admissible in the full structured model. For instance, at first glance, nothing appears to prohibit the system evolving to the physically infeasible case of $\mu_0=0$ whilst $\mu_1>0$. This is, in fact, a special case of the general class of physically unrealistic states characterised by $\mu_1 > N\mu_0$. However, 
it turns out that 
the derived system of moments does respect this constraint: if $\mu_1=N\mu_0$, then $\mathrm{d}\mu_1/\mathrm{d}t = N\mathrm{d}\mu_0/\mathrm{d}t$, so that a trajectory initially satisfying $\mu_1 \leq N\mu_0$ will do so for all time. Hence, feasible states remain feasible for all time. This compatibility condition, along with the analogous constraints for higher order moments, will be present in all derived moment equations and serves as a simple way of verifying the correctness of computed closures. We can see examples of this in \cref{fig: moments of model A,fig: exponential phase portraits}, where the shaded regions are infeasible and unreachable.

\subsubsection{Nonlinear state dependence}
As a final example, we now seek closures of the most complex model of \cref{sec: examples: complex}, defined by \cref{eq: complex example: model D}. The dynamics of each subpopulation depend explicitly on the first moment $\mu_1$, so that any closures must be at least of order $1$, with
\begin{equation}
    n(j) = \frac{2}{N}(N-j)^2 - \left(1 + \frac{N-j}{N}\right)\mu_1\,.
\end{equation}
We do not expect to be able to close this system, as $\deg{n}=2$. In particular, we expect to be unable to simultaneously eliminate higher order moments from the evolution equations of $\mu_{K-1}$ and $\mu_K$ for any $K$.

To verify this prediction, we now seek a closure of order $K=2$. We impose the degree and coefficient constraints of \cref{eq: general coefficient constraint,eq: degree constraint}, so that $\deg{w}=3$ and its coefficients satisfy
\begin{subequations}
\begin{align}
    \frac{\mu_1}{N} - 4 + 2\omega_2 + \omega_3 &= 0\,,\\
    \frac{2}{N} + 2\omega_3 &= 0\,,
\end{align}
\end{subequations}
where we have computed $\phi_{2,2} = 2\omega_2+\omega_3$ and $\phi_{2,3} = 2\omega_3$ from \cref{eq: convolution}. Solving for the coefficients of $w$ yields fluxes of the general form
\begin{equation}
    w(j) = -\frac{1}{N}j^3 + \left(2 - \frac{\mu_1 - 1}{2N}\right)j^2 + \omega_1j + \omega_0\,,
\end{equation}
where $\omega_1$ and $\omega_0$ are yet to be determined. This flux rate depends explicitly on the moment $\mu_1$, a feature that has been inherited from the growth dynamics that will not prove problematic. Imposing $w(N)=0$ then reduces this to a one-dimensional family of admissible fluxes, further constrained by the requirement that $w(j)\geq0$ for $j\in\{0,\ldots,N-1\}$, though we omit these details here for brevity.

The imposed conditions are necessary and sufficient for removing higher order moments from the evolution equation for $\mu_2$. To verify this, we compute the time evolution of moments $\mu_0$, $\mu_1$, and $\mu_2$ as
\begin{subequations}
\begin{align}
    \diff{\mu_0}{t} &= \frac{2}{N}\mu_2 + \left(\frac{\mu_1}{N} - 4\right)\mu_1 + 2(N-\mu_1)\mu_0 \,,\\
    \diff{\mu_1}{t} &= \frac{1}{N}\mu_3 + \left(\frac{\mu_1+1}{2N} - 2\right)\mu_2 + 2(N-\mu_1+\omega_1)\mu_1 + \omega_0\mu_0 \,,\\
    \diff{\mu_2}{t} &= \left(2[1+\omega_1 + N-\mu_1]- \frac{\mu_1-1}{2N}\right)\mu_2 + (2\omega_0 + \omega_1)\mu_1 + \omega_0\mu_0 \,.
\end{align}
\end{subequations}
We note that, as expected, the evolution of $\mu_2$ depends only on $\mu_0$, $\mu_1$, and $\mu_2$. However, the evolution of $\mu_1$ irreparably depends on $\mu_3$, with coefficient $\nu_2 + \omega_3 \neq 0$. Hence, we verify that exact moment closure is unattainable at order 2 in this system, driven by the nonlinear dependence of the growth kinetics on state. More generally, the coefficient of $\mu_{K+1}$ in the evolution equation for $\mu_{K-1}$ will be $\nu_2/K\neq0$, so that closure is impossible at all orders in this system. Notably, this approaches zero for large $K$, so one might expect in future work to find that approximate closures of reasonable accuracy can be found at high order when $\deg{n}=2$.

\section{Identifying mechanism in essentially unstructured systems}
\label{sec: state-preserving vs state-resetting}
To conclude, we highlight the utility of structured population models for investigating the mechanisms underlying population growth. In \cref{sec: formulation}, specifically \cref{eq: reproduction rate}, we noted the existence of two plausible mechanisms of growth in structured populations: state-preserving growth (offspring are born into the compartment containing their parent) and state-resetting growth (offspring enter $u_0$). In any given experimental system, it may not be clear which of these mechanisms is at play. However, if we restrict ourselves to essentially unstructured systems, where one might think that modelling structure is superfluous, we will show that structured models can allow us to distinguish between these two mechanisms using only population-level information.

Consider two essentially unstructured populations that can be modelled via \cref{eq: general model} with the restrictions of \cref{sec: essentially unstructured analysis}. Informed by the analysis of that section, we take $w(j)$ to be linear in $j$, so that the system of moments closes at all orders. Their evolution can be summarised as
\begin{equation}
    \diff{u_j}{t} = \rho u_j + \delta_{0j}r\sum_i  u_i + \underbrace{\left[J_j - du_j - eu_j\sum_iu_i \right]}_{L_j} \,,
\end{equation}
where here $\rho$, $r$, $d$ and $e$ are constant rates of growth or death and we have grouped together all the terms unrelated to reproduction into a single term $L_j$. Notably, by the above constraints on $w(j)$, we have that $\sum_jL_j = -(d+e\mu_0)\mu_0$ and that $\sum_jjL_j=f(\mu_0,\mu_1)$ is a function of $\mu_0$ and $\mu_1$ only. 

Suppose that the two populations of interest are identical save for their mechanism of growth: one performs only state-preserving growth, whilst the other performs only state-resetting growth. Hence, writing $u_j^P$ and $u_j^R$ for the state-preserving and state-resetting populations, respectively, they satisfy the following evolution equations:
\begin{equation}\label{eq: almost identical populations: P}
    \diff{u_j^P}{t} = \rho u_j^P + L_j^P
\end{equation}
and
\begin{equation}\label{eq: almost identical populations: R}
    \diff{u_j^R}{t} = \delta_{0j}r\sum_i u_i^R + L_j^R\,.
\end{equation}
Here and throughout, superscripts of $P$ and $R$ correspond to state-preserving and state-resetting analogues, respectively. Note that, in order for only the mechanism (rather than the rate) of growth to differ between populations, we must have $\rho = r$.

Together, these constraints have a practically unfortunate, if unsurprising, consequence: were we to measure the total size of each population, the two would be indistinguishable. This is apparent when individually summing \cref{eq: almost identical populations: P,eq: almost identical populations: R} over $j$ to yield evolution equations for $\mu_0^P$ and $\mu_0^R$:
\begin{subequations}
\begin{align}
    \diff{\mu_0^P}{t} &= \rho\mu_0^P - (d+e\mu_0^P)\mu_0^P\,,\\
    \diff{\mu_0^R}{t} &= r\mu_0^R - (d+e\mu_0^R)\mu_0^R\,,
\end{align}
\end{subequations}
where $\rho = r$ ensures that both populations will evolve in the same way from equal initial conditions.

Thus far, it looks like structure has afforded little other than notational complexity. However, computing the evolution equation for the first moment of the populations reveals the benefit of explicitly modelling structure. Multiplying by $j$ and summing gives
\begin{subequations}
\begin{align}
    \diff{\mu_1^P}{t} &= \rho\mu_1^P + f(\mu_0^P,\mu_1^P) - (d+e\mu_0^P)\mu_1^P\,,\\
    \diff{\mu_1^R}{t} &= f(\mu_0^R,\mu_1^R) - (d+e\mu_0^R)\mu_1^R\,,
\end{align}
\end{subequations}
where the terms involving $f$ arise from summing $jL_j$. From the form of these equations, we see that only state-preserving reproduction has a direct and immediate impact on the evolution of the first moment of the state index. In contrast, whilst state-resetting reproduction naturally impacts the whole dynamics, its effects are only via the eventual evolution $\mu_0$ and $\mu_1$. Concretely, given equal starting conditions, the initial rate of change of $\mu_1^P$ depends strongly on $\rho = r$, whilst the initial evolution of $\mu_1^R$ is independent of the growth rate. 

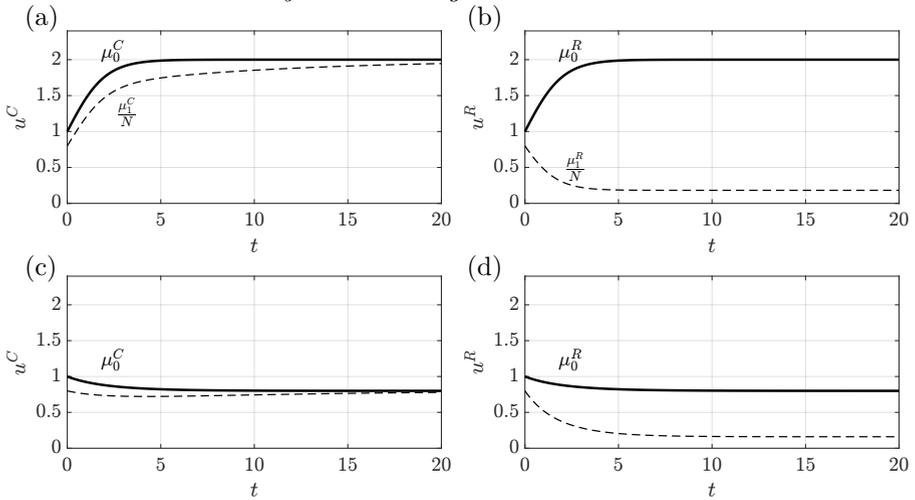
\begin{figure}
    \centering
    \begin{overpic}[width=\textwidth]{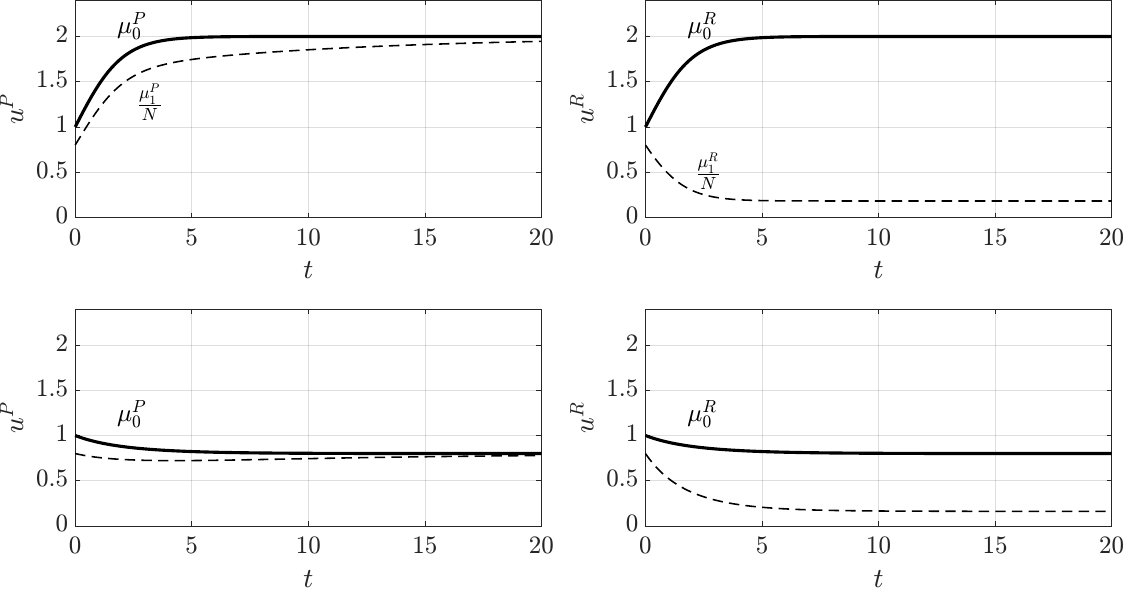}
\put(2,53){(a)}
\put(51,53){(b)}
\put(2,25){(c)}
\put(51,25){(d)}
\end{overpic}
    \caption{The dynamics of four essentially unstructured models, highlighting the information that structure can provide even in seemingly unstructured systems. In (a) and (b) we plot the evolution of populations $u^P$ and $u^R$, which grow via state-preserving and state-resetting reproduction, respectively, with equal parameters. The different mechanisms of growth lead to different first moments $\mu_1^P$ and $\mu_1^R$ (shown dashed) but equal total populations (solid black curves). In (c) and (d) we reduce the growth rate by 60\% from (a) and (b), leading to a reduction in overall population size. Notably, the evolution of $\mu_1^P$ is markedly different to that in (a). In contrast, the initial evolution of $\mu_1^R$ remains approximately unchanged between (b) and (d), distinguishing state-resetting growth from state-preserving growth through their first moments. Here, we have taken $N=10$, $g=1$, $w(j)=(N-j)/10$ and $K=2$, with $\rho=r=1$ in (a) and (b).}
    \label{fig: moments revealing mechanism}
\end{figure}

This is exemplified in \cref{fig: moments revealing mechanism}, where we compare the evolution of two such populations and their response to a change in growth rate. In \cref{fig: moments revealing mechanism}a,b we plot the evolution of a state-preserving and state-resetting population, respectively, with the scaled first moments $\mu_1^P/N$ and $\mu_1^R/N$ shown dashed. Notably, though the first moments are different, the evolution of the total population is the same in each case. When changing the growth rate of both populations (panels c and d), we note that the evolution of $\mu_1^P$ (shown dashed in \cref{fig: moments revealing mechanism}c) is drastically impacted by the parameter change, with its derivative even switching sign, whilst the evolution of $\mu_1^R$ in \cref{fig: moments revealing mechanism}d appears insensitive to the change in parameter, at least at short times.

Overall, this highlights that, were one able to modify the growth rate in a controlled system, measurements of only $\mu_1$ would allow for the identification of the mechanism of growth, whilst measuring only $\mu_0$ (which may seem more natural) would not. In particular, an observation of $\mu_1$ being sensitive to the change in growth rate would signify the presence of state-preserving growth, whilst insensitivity during initial evolution would suggest state-resetting growth. Thus, even in essentially unstructured systems, modelling and measuring what little structure is present can illuminate otherwise unidentifiable details of the dynamics.

\section{Discussion}

In this introductory article, we have explored the dynamics of structured populations that can be described by generalisations of the classical logistic model of growth. Perhaps surprisingly, even generalising the principles of logistic growth led to some ambiguity in the mechanism of reproduction: is the structure variable preserved during reproduction, or is it changed in the process? Here we considered just two possibilities, either preserving or resetting the state of newly spawned population, and found in \cref{sec: state-preserving vs state-resetting} that the study of moment closures can distinguish these two mechanisms in simply structured populations using aggregated information alone. In future, a more complete investigation into the possible effects of reproduction on the structure variable is warranted, including an exploration of average birth and death rates, especially given the prospect of using population-level information to discern details of mechanism.

A key focus of this study has been seeking out conditions for exact moment closures, which we demonstrated to be very restrictive. Though the relative rarity of closures may not be too surprising, it does highlight how remarkable it is that closures have been found in previous works \citep{Chambers2023AMacrophages,Ford2019EfferocytosisAccumulationinsidemacrophagepopulations}, especially as these systems do not appear to have been contrived with the intent of generating exact closures. The model of \citep{Chambers2023AMacrophages} is readily cast in the framework of this manuscript and constitutes an essentially unstructured system with $\deg{w}=1$. Hence, by the analysis of \cref{sec: essentially unstructured analysis}, it admits closures at all orders. Two of these were in fact noted by \citep{Chambers2023AMacrophages}: the zeroth-order closure and a non-trivial closure at order $K=1$. \citep{Ford2019EfferocytosisAccumulationinsidemacrophagepopulations} considered a range of models, one of which was also essentially unstructured. Unfortunately, a complete analysis of this model lies just outside the scope of this study, as the model includes a convolution of the form $\sum_j u_ju_{N-j}$ in its growth kinetics, though it also represents an essentially unstructured system and thus closes at zeroth order. 

More generally, looking towards considering systems where exact closures are not realisable, it is not clear how significant the errors associated with slightly violating the derived closure conditions (e.g. by perturbing a coefficient constraint) might be. This prompts the future pursuit of systematically constructing approximate moment closures, perhaps drawing on the wealth of literature in the active matter and stochastic systems communities \citep{Saintillan2013ActiveModels,Cintra1995OrthotropicOrientation,Kuehn2016MomentReview, Gillespie2009Moment-closureModels}, for instance. The existing progress in these other fields also motivates a future study of the impacts of taking the continuum limit ($N\rightarrow\infty$) of the discretely structured systems that we've explored in this work. A priori, it is not clear if the conclusions reached in this study will carry over to the continuum limit. Indeed, with reference to the continuum limit obtained in \cite{Chambers2023AMacrophages}, we expect there to be significant nuance in how this limit should be taken, which may depend on the biological context in question.

Having only considered closures of a single structured population, there are a multitude of biologically relevant extensions that could be considered. Perhaps the most natural generalisation is to account for multiple interacting species that are themselves structured, exploring if and how the interactions between the species impacts on the form and existence of exact closures. A further extension that aligns closely with biological and medical applications is the consideration of alternative mechanisms of reproduction \citep{Savill2003MathematicalEpidermis,Komarova2013PrinciplesLineages,Yang2015TheHomeostasis}, or the impacts of a time-dependent external factor such as a drug on multicellular tumour growth. Whilst evaluating the effects of state-independent factors may be relatively straightforward, the arguably more intriguing possibility for state-dependent effects raises non-trivial mathematical questions and promises to be a topic for significant future study. Additionally, an attempt to move beyond polynomial kinetic functions is a promising but challenging direction for future work, as our analysis has been very sensitive to our choice of function space. Initial explorations suggest that this is expected to require case-dependent alternatives to $\mu_k$ that behave suitably under summation.

In summary, we have taken a classical model of population growth and begun to explore the exact study of discretely structured analogues, with a particular focus on moment closures. Restricting our exploration to a class of models with a polynomial dependence on the structure variable, we have identified necessary and sufficient conditions for systems to admit exact closures. In doing so, we have seen how restrictive the desire for exact closures can be, whilst also finding that some freedom remains once any constraints have been satisfied. Additionally, we have highlighted the potential benefits of studying structure even in simply structured populations, with signatures present in structured models able to yield mechanistic insight that might otherwise be obfuscated by unstructured approaches.

\section*{Acknowledgements and competing interests}
BJW is supported by the Royal Commission for the Exhibition of 1851. The authors have no competing interests to declare.

\section*{Data availability}
All computer code used to generate figures in this work is available at \url{https://github.com/Mar5bar/structured-logistic-models}.

\begin{appendices}
\section{Flux in both directions}\label[app]{app: both fluxes}
The above analysis can be modified in a straightforward way to include fluxes in the direction of decreasing $j$. Explicitly, we introduce the flux $\tilde{J} = \tilde{w}(j+1)u_{j+1} - \tilde{w}(j)u_{j}$, where $\tilde{w}(j)=\sum_{\beta = 0}^{\deg{\tilde{w}}}\tilde{\omega}_{\beta} j^\beta$ and $\tilde{\omega}_{\beta}$ are the coefficients of the flux, analogous to $\omega_{\beta}$ in the other direction and subject to similar conditions at the edges of the domain ($\tilde{w}(0) = 0$ and $\tilde{w}(N+1) = 0$ for convenience).

Adding this flux term to the general analysis of \cref{sec: general conditions} modifies \cref{eq: unidirectional moment evolution} to 
\begin{equation}
    \diff{\mu_k}{t} = \sum_j \underbrace{\left\{j^k n(j) + \left[(j+1)^k - j^k\right]w(j) + \left[(j-1)^k - j^k\right]\tilde{w}(j)\right\}}_{\tilde{p}(j)}u_j\,,
\end{equation}
where $\tilde{p}(j)$ can be written as
\begin{equation}
    \tilde{p}(j) = \sum_{\alpha = 0}^{\deg{n}} \nu_{\alpha} j^{k + \alpha} + \sum_{\beta = 0}^{k-1 + \deg{w}}\phi_{k,\beta} j^{\beta} + \sum_{\beta = 0}^{k-1 + \deg{\tilde{w}}}\tilde{\phi}_{k,\beta} j^{\beta}
\end{equation}
and the $k$-dependent coefficients $\tilde{\phi}_{k,\beta}$ are defined via the convolution
\begin{equation}
    \tilde{\phi}_{k,\beta} = \sum_{\gamma_1 + \gamma_2=\beta} \tilde{\omega}_{\gamma_1}\binom{k}{\gamma_2}(-1)^{\gamma_2}
\end{equation}
for $\gamma_1\in\{0,\ldots,\deg{\tilde{w}}\}$ and $\gamma_2\in\{0,\ldots,k-1\}$. The moment evolution equation is then
\begin{equation}
    \diff{\mu_k}{t} = \sum_{\alpha = 0}^{\deg{n}} \nu_{\alpha} \mu_{k + \alpha} + \sum_{\beta = 0}^{k-1 + \deg{w}}\phi_{k,\beta} \mu_{\beta} + \sum_{\beta = 0}^{k-1 + \deg{\tilde{w}}}\tilde{\phi}_{k,\beta} \mu_{\beta}\,.
\end{equation}
The resulting conditions for closure can then be derived in exactly the same manner as in \cref{sec: general conditions}.  Once again, the results differ depending on the complexity of the growth terms. If $\deg{n} = 1$, closures can be found at any order through appropriate choices of the flux terms, but now with additional degrees of freedom afforded by the presence of $\tilde{\omega}$. For instance, if $w$ and $\tilde{w}$ are of the same degree, then the necessary and sufficient conditions for closure at order $K$ amount to the degree constraint
\begin{equation}
    \deg{n} = \deg{w} - 1 = \deg{\tilde{w}} - 1\,,
\end{equation}
and the single coefficient constraint
\begin{equation}
    \nu_{\alpha} + K\left[w_{\deg{w}} - \tilde{w}_{\deg{\tilde{w}}}\right] = 0\,.
\end{equation}
Combined with the requirements that $w(N) = 0$ and $\tilde{w}(0) = 0$, this amounts to 3 constraints on 6 coefficients (as both $w$ and $\tilde{w}$ are quadratics in $j$), so that significant freedom is afforded by having bidirectional fluxes. In contrast, for $\deg{n}>1$, we again encounter the problem illustrated in \cref{eq: subleading moment problem}, in which the dependence on $\mu_{K-1+\deg{n}}$ cannot be eliminated in the equation for $\mu_{K-1}$. Hence, closures are never possible for $\deg{n}>1$, even with bidirectional transport.

\end{appendices}

\bibliography{references.bib}

\end{document}